\documentclass[conference]{IEEEtran}

\usepackage{booktabs}
\usepackage{cite}
\usepackage{amsmath,amssymb,amsfonts}
\usepackage{graphicx}
\usepackage{textcomp}
\usepackage{xcolor}
\def\BibTeX{{\rm B\kern-.05em{\sc i\kern-.025em b}\kern-.08em
    T\kern-.1667em\lower.7ex\hbox{E}\kern-.125emX}}
\usepackage{array}
\usepackage{tabularx}
\usepackage{braket}
\usepackage{hyperref}

\begin{document}

\title{Benchmarking Quantum and Classical Machine Learning Models on Oncological Data}

\author{\IEEEauthorblockN{Sydney Leither}
\IEEEauthorblockA{\textit{Cascade Quantum Inc}\\
East Lansing, USA \\
sleither@cascadequantum.com}
\and
\IEEEauthorblockN{Thomas Lubinski}
\IEEEauthorblockA{\textit{Cascade Quantum Inc}\\
San Rafael, USA \\
tlubinski@cascadequantum.com}
\and
\IEEEauthorblockN{Michael Kubal}
\IEEEauthorblockA{\textit{Cascade Quantum Inc}\\
Chicago, USA \\
mkubal@cascadequantum.com}
\and
\IEEEauthorblockN{Sonika Johri}
\IEEEauthorblockA{\textit{Cascade Quantum Inc}\\
Cupertino, USA \\
sjohri@cascadequantum.com}
}

\maketitle

\begin{abstract}
Machine learning is being increasingly used for the detection, diagnosis, and treatment of cancer. However, models often struggle with biological data due to high dimensionality, limited sample diversity, and complex feature interactions. Recent works have investigated the potential for quantum machine learning models to exhibit improved performance over classical models on this kind of complex data, but have often lacked rigorous empirical evaluation of quantum advantage. In this work, we develop a methodology for fair benchmarking of quantum and classical machine learning models, based on the Red Cedar quantum machine learning and resource estimation framework and AutoML-optimized classical neural networks. We assess the potential for quantum advantage in machine learning across tabular, omics, and spatial oncological datasets drawn from the existing quantum machine learning literature, with a range of preprocessing methods, and find no evidence of quantum advantage. Our results suggest that the field should prioritize analyzing higher-dimensional, more biologically realistic datasets to make meaningful progress toward practical quantum advantage in oncological classification problems.
\end{abstract}


\section{Introduction}
Biological systems exhibit cross-scale, nonlinear interactions that make biological data especially difficult to model \cite{peters2004cross}. In particular, the biological scales relevant to oncological processes include molecular, cellular, tissue, patient, and societal \cite{kalfon2025towards}. Each of these scales emerges from interactions and processes from the preceding scales, and each has its own data modalities that researchers and practitioners use to understand and treat cancer. Molecular data modalities, for example, include genomic, transcriptomic, proteomic, metabolomic, and epigenomic \cite{gupta2024chapter}. Datasets containing some or all of these ``-omic'' data types are referred to as multi-omics data, and extracting the information from these data is an on-going challenge \cite{acharya2024comprehensive}. Yet multi-omics captures only one scale of oncological data, and both combining and modeling multiple scales of data is another challenge \cite{kalfon2025towards}.

One of the main drivers for modeling oncological data is for the detection, diagnosis, and treatment of cancer \cite{craddock2022evaluation}. Increasingly, researchers are attempting to use machine learning to learn patterns in existing data and then predict or classify aspects of new data, such as presence of cancer in a patient \cite{aftab2025ai}. The particularities of both data sampling methods and machine learning models create challenges in achieving the goal of making useful insights from oncological data \cite{lotter2024artificial}. One challenge is the limited number of samples: medical data is protected and thus not commonly accessible to outside researchers. Even for those with access to data, there are limitations in the number and diversity of samples, which leads to models that cannot be applied widely \cite{tasci2022bias}. Another challenge is the quality of the data once it is sampled --- missing and erroneous data are common \cite{flores2023missing} and must be addressed before downstream machine learning \cite{song2023learning}. Furthermore, even with significant data present and cleaned, the complexity of the high number of features common in oncological data leads to the ``curse of dimensionality'' \cite{donoho2000high}, which is a well-known issue that plagues the performance of traditional machine learning models \cite{bubeck2023universal}.

Quantum computing offers a promising approach to tackle the complexity inherent in oncological data \cite{ramesh2024quantum}. Quantum computing leverages quantum mechanical phenomena to perform computations that are exponentially costly in classical computing. The properties of superposition, entanglement, and interference offer an approach for more parallelizable and scalable computation, but the loss of information from measurement and decoherence of physical qubits demand the development of specialized algorithms to realize quantum advantage. Quantum advantage is when a quantum algorithm shows superior efficiency, cost-effectiveness, or accuracy than an analogous classical algorithm \cite{lanes2025framework}. Significant attention has been given to theoretical quantum advantage in machine learning \cite{cerezo2022challenges}, but evidence of practical quantum advantage is debated \cite{bowles2024better}.

Here, we consider a form of quantum machine learning in which we optimize a model composed of parameterized operations for a given learning task, with all or part of the model able to be executed on a quantum computer \cite{du2025quantum}. Researchers are currently exploring a number of settings in which this approach could improve learning on oncological data  \cite{flother2025quantum}. One study suggests that quantum models are expected to generalize better on fewer samples of data \cite{caro2022generalization}, which may lead to better predictions on oncological data, which often has a limited diversity of samples \cite{tasci2022bias}. Quantum methods are also expected to have advantages over classical machine learning in model size and performance on the kinds of high-dimensional data common in oncology \cite{abbas2021power, molteni2026quantum}. While there are many claims of expected advantages for quantum machine learning, it is difficult to rigorously prove these claims as machine learning problems are notoriously hard to study theoretically \cite{schuld2022quantum}. A standard and rigorous framework for evaluating quantum advantage in machine learning is needed \cite{huang2021power, bowles2024better}, especially for validating claims of promise for quantum computing in the life sciences \cite{maurizio2025quantum, gupta2025systematic}.

In this paper, we investigate the potential for quantum advantage in oncological classification tasks across multiple data modalities and preprocessing strategies. Prior studies have not disentangled whether the performance of quantum models is limited by optimization challenges, such as convergence to local minima or initialization in barren plateaus, by the intrinsic expressivity of the model, or the generalization abilities of a particular quantum architecture or data encoding scheme. A similar issue arises in classical baselines, where models are often trained with fixed optimizer hyperparameters, leaving open the question of whether their best achievable performance has been realized.

To address these limitations, we adopt recently proposed techniques that enable quantum models to be trained to arbitrarily high training accuracy without encountering barren plateaus or expressivity constraints \cite{johri2025bit, leither2025qubits}. These methods, implemented within the (in-development) Red Cedar framework, allow us to decouple optimization effects from model capacity. For classical models, we employ neural networks optimized via the TPOT2 AutoML library \cite{ribeiro2024tpot2}, using explicit stopping criteria. As a result, training performance is no longer a limiting factor for either paradigm, allowing us to focus the analysis on generalization performance as measured on held-out test data.

We then evaluate both model classes, as well as a theoretical estimate based on resource estimation, on oncological classification problems spanning tabular, omics, and spatial datasets, under a range of preprocessing methods. Across these settings, we find no evidence of quantum advantage on commonly used oncological benchmark datasets in the quantum machine learning literature. These results suggest that demonstrating such an advantage may require moving beyond current benchmarks toward more complex and higher-dimensional data. More broadly, our work establishes a more rigorous framework for assessing quantum advantage by systematically analyzing performance scaling with model size for both quantum and classical models, while reporting appropriate statistical measures \cite{gupta2025systematic}.

\section{Methods}

\begin{figure*}
\centering
\includegraphics[width=\linewidth]{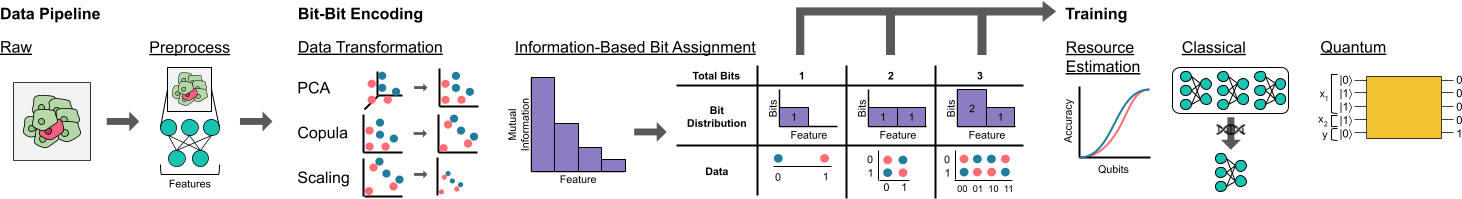}
\caption{The data processing steps for the experiments in the paper. First, the ``data pipeline'' step optionally applies feature selection or feature reduction preprocessing to reduce noise in the dataset. Then, the ``bit-bit encoding'' step makes the dataset quantum-ready by applying classical data transformations to further reduce noise before discretizing the data into bits using an information-based bit assignment. Finally, the ``training'' steps trains models on the dataset discretized into different numbers of bits.}
\label{fig:fig1}
\end{figure*}

\subsection{Data}
We investigate the simple Wisconsin Diagnostic Breast Cancer (WDBC) dataset \cite{wdbc} and multiple kinds of oncological omics and spatial data. We choose omics and spatial data modalities specifically due to claims of a higher potential for quantum machine learning advantage with those modalities \cite{ramesh2024quantum, flother2025quantum}. We restrict ourselves to oncological datasets that have appeared previously in the quantum machine learning literature and are appropriate for multiclass supervised machine learning. The spatial datasets are processed such that image-specific models, such as convolutional neural networks, are not necessary.

\subsubsection{Wisconsin Diagnostic Breast Cancer} \label{sec:wdbc}
We place the WDBC dataset \cite{wdbc} in its own category as it is a uniquely simple benchmark dataset common in the oncological machine learning literature. The WDBC dataset contains summary statistics about the size, shape, and texture of cell nuclei present in images from a fine needle aspirate slide of malignant and benign breast tumors \cite{street1993nuclear}. WDBC is the most common oncological dataset in the quantum machine learning literature, with the majority using quantum support vector machines as their learning algorithm \cite{vashisth2021design, moradi2022clinical, shan2022demonstration, premanand2023quantum, wang2024novel, desai2024comparison, bhuvaneshwari2025integrating}. We note that quantum support vector machines have unfavorable scaling of inference time with the number of training samples in the dataset. Thus, we focus on quantum variational circuits in this work, similar to other works using this dataset \cite{premanand2023quantum, desai2024comparison, kundu2025harnessing}.

\subsubsection{Omics}
Omics data, which characterizes and quantifies biological molecules, forms the basis of precision medicine and understanding the mechanisms of cancer. Omics datasets are thought to be promising candidates for quantum machine learning advantage due to their propensity to exhibit the ``curse of dimensionality'' and complex, higher-order interactions \cite{ramesh2024quantum, flother2025quantum}. We use multi-omics data from The Cancer Genome Atlas (TCGA), a repository of oncological omics datasets widely used by machine learning researchers. TCGA contains genomic, transcriptomic, and epigenomic data of multiple kinds of cancer. Previous quantum work classified cancer types and discovered potential biomarkers using TCGA data \cite{li2021quantum, ghobadi2024potential, nguyen2024biomarker, kaveh2025investigating, saggi2026multi}. In this work, we use the MLOmics library \cite{yang2025mlomics} to access cleaned and processed TCGA data for cancer subtype classification. We investigate the potential for quantum advantage on the three original-feature ``golden-standard'' subtype classification datasets with the most samples within MLOmics: GS-BRCA (breast invasive carcinoma), GS-COAD (colon adenocarcinoma), GS-OV (ovarian serous cystadenocarcinoma). We use the genomic mRNA gene expression data from each dataset as the features; we leave the exploration of the potential for quantum advantage on these datasets with integrated multi-omics to future work.

\subsubsection{Spatial}
The interpretation of imaging data, such as X-ray, MRI, or histopathological images, forms the basis of cancer diagnostics. Many studies have applied quantum machine learning to medical imaging data \cite{wei2023quantum, kaveh2025investigating, radhi2025quantum}. Here, we investigate the potential for quantum machine learning advantage on the BreastMNIST, PathMNIST, and DermaMNIST oncological datasets from MedMNIST \cite{yang2023medmnist}, a standardized collection of benchmark medical imaging datasets. BreastMNIST contains breast ultrasound images labeled as cancerous or not, PathMNIST contains colorectal cancer histology slides labeled with tissue type, and DermaMNIST contains images of pigmented skin lesions labeled with presenting disease. Multiple previous works have classified MedMNIST datasets using quantum approaches \cite{qiu2023universal, landman2022quantum, krishnakumar2025extreme, rahim2025hybird}. Note that we remove two colliding samples, \textit{i.e.} samples with the same image but different class labels, from both the PathMNIST and BreastMNIST datasets.

\subsection{Preprocessing} \label{sec:preprocessing}
An important aspect of machine learning is preprocessing a model's input data before training the model in order to optimize performance. Preprocessing steps may include any of the following: data cleaning, normalization, encoding, transformation, and feature reduction. Data preprocessing is especially important for quantum machine learning due to the limited capacity of current quantum hardware. Data must be preprocessed such that the greatest amount of relevant information from the features can fit into the number of qubits available on the quantum hardware. We test the effect of different data modality-specific preprocessing algorithms present in the existing literature on the potential for quantum advantage. The preprocessing methods we employ are all purely classical and scale polynomially in the size of the dataset.

For the WDBC dataset and the omics datasets, we compare results without preprocessing and with random forest feature selection preprocessing. We choose random forest feature selection due to its prevalence in existing work on quantum machine learning for omics data \cite{mondal2025quantum, saggi2026multi}. We implement random forest feature selection by training decision tree classifiers from scikit-learn \cite{pedregosa2011scikit} on 100-sample batches of the training data and then maintaining all features that have an average importance score greater than zero.

For the spatial datasets, we compare results when using flattened image pixel values and using the output vector from passing images through ResNet18 \cite{he2016deep}. ResNet18 is an 18-layer convolutional neural network with residual learning. We use a ResNet18 model pre-trained on ImageNet \cite{deng2009imagenet}, which means the model has already learned some of the complex structures seen in a wide variety of images. We preprocess our spatial data using the pre-trained ResNet18 by passing each image through the model and using the last layer of weights as the feature vector, resulting in 512 features per image post-preprocessing. While the most common approach to image classification in the quantum machine learning literature is using convolutional layers \cite{farooq2025systematic}, that requires the development of a separate model, which we leave to future work. Previous work has also explored flattening images for quantum machine learning and transforming images into feature vectors using a pre-trained model such as ResNet18 \cite{farooq2025systematic}.

\subsection{Quantum machine learning}
\subsubsection{Data encoding} \label{sec:encoding}
One of the biggest current algorithmic challenges to realizing the promise of quantum machine learning is the encoding of classical data into a quantum representation \cite{cerezo2022challenges, maurizio2025quantum}. The most common data encoding approaches are angle encoding, where data are encoded into the rotation angle of qubits, and amplitude encoding, where data are embedded into the amplitudes of a quantum state \cite{weigold2021encoding}. Quantum machine learning algorithms utilizing these encoding methods have limited expressivity, \textit{i.e.} they limit the class of functions the algorithm can learn when the number of times the data are uploaded is finite \cite{schuld2021effect, wang2025limitations, johri2025bit}, which limits the maximum attainable performance for learning problems, especially on complex datasets.

Quantum machine learning algorithms utilizing bit-bit encoding \cite{johri2025bit} have the property of universal approximation, meaning they achieve full expressivity with a single uploading of the encoded data, and could theoretically achieve $100\%$ training accuracy on any arbitrary dataset \cite{leither2025qubits}. Therefore, we employ bit-bit encoding in our experiments. Bit-bit encoding classically encodes each sample in a dataset into bits, which then correspond to initialization and readout of qubits in the computational basis. The input is loaded into a register of $Q_n$ ``data qubits'' and output is read from a separate register of $Q_c$ ``label qubits''. The goal of bit-bit encoding is to compress the dataset into a user-specified number of bits while retaining the most classification-relevant information. 

Figure~\ref{fig:fig1} visualizes the steps of bit-bit encoding. First, preprocessing is optionally applied to a raw dataset to reduce noise in the dataset for better downstream model performance. Then bit-bit encoding has two stages: data transformation and information-based bit assignment (discretization). The data transformation step applies z-score normalization, PCA dimensionality reduction, Copula transform, and min-max normalization to the data. In this work, we set the number of components (features) after PCA dimensionality reduction to be the number of components at which $90\%$ of the variance is explained. The information based-bit assignment step calculates the mutual information between each retained component and the class. Then, bits are allocated to the components such that the total number of bits sums to a user-specified amount and each component has a number of bits proportional to its mutual information ratio. For example, if component zero contains $50\%$ of the total mutual information shared between each component and the class, then $50\%$ of the total bits will be allocated to that component. Each value $x$ in each component is discretized into the $b$ bits assigned to the component with the formula $\lfloor x2^{b}\rfloor$. For example, a value of $0.25$ within a component being discretized into three bits becomes $\lfloor 0.25 \cdot2^3 \rfloor = 2 \rightarrow 010$. For loading the bit-bit encoded data as basis states of a quantum model, the class labels must also be encoded with $\lceil \log_2\text{(number of classes)} \rceil$ bits.

\subsubsection{Resource estimation} \label{sec:re}
The universal approximation property of bit-bit encoding allows it to form the foundation of a resource estimation framework \cite{leither2025qubits}. With this, we can calculate the number of qubits required to encode a dataset to a desired degree of accuracy. The maximum theoretical accuracy a model could achieve on a dataset depends on the number of bits allocated to the dataset. With bit-bit encoding, datasets are discretized to different levels of precision depending on the bits allocated. When a dataset is discretized, samples that were unique before discretization can map to the same bitstring. If those samples have different class labels, we call this a collision. The theoretical accuracy is limited by the number of collisions in the discretized dataset, as the information needed to distinguish the classes has been lost. The theoretical accuracy can thus be computed from the frequency of collisions.

The testing accuracy is calculated using collisions in the overlap between train and test, which is when an encoded sample in the test set has a different class label than the majority of the same encoded samples in the train set. Note that the overlap, by definition, cannot account for test samples that are not in the training set, so when a test sample is not in the training set it counts as correctly ``classified'' for the theoretical accuracy metric. This means the theoretical test accuracy is realistically too optimistic and accounting for this is left for future work.

The number of bits at which it is theoretically possible to reach $100\%$ train and test accuracy is called $Q_{dataset}(1.0)$. This metric is calculated by encoding a dataset to an increasing number of bits until there are no more collisions or overlap. If a dataset's $Q_{dataset}(1.0)$ is above $50$, it suggests the classification problem has a potential for quantum advantage, as that is around when classical simulation capabilities plateau \cite{decross2025computational}. Note that $Q_{dataset}(1.0)$ also contains the number of qubits required to encode the class labels.

\subsubsection{Model Architecture}
The quantum model uses a layered bipartite architecture, where a single layer consists of parameterized entangling nodes between each unique combination of label qubit and data qubit. Entangling nodes consist of single qubit Euler rotations followed by a Heisenberg-type unitary acting on both qubits \cite{johri2025bit}. In other words, each entangling node corresponds to a parameterized rotation of each qubit on its Bloch sphere before a parameterized entangling operation. There are then a total of $\frac{9}{2}Q_nQ_c$ parameters (angles of rotation) per layer to optimize during training. A parameterized Euler unitary is added to each label qubit before it is measured, thus adding a final $3Q_c$ parameters. Since the parameterized gates are from a universal gate set, and the connectivity structure enables entanglement between any pair of qubits, this model architecture alongside bit-bit data encoding leads to universal approximation as the number of layers increases.

\subsubsection{Model Training}
We train a quantum model for each preprocessing method, dataset, and train-test split. A major challenge in quantum machine learning is the trainability of quantum models \cite{maurizio2025quantum, cerezo2022challenges}. Quantum model loss function landscapes can become exponentially flat as the number of qubits increase, a phenomenon known as a barren plateau \cite{mcclean2018barren}. Techniques to avoid barren plateaus include using shallower or highly-structured model architectures, which reduces the potential for quantum advantage \cite{cerezo2025provable}. To avoid this, the Red Cedar quantum machine learning framework \cite{johri2025bit} employs a novel sub-net initialization approach. Sub-net initialization uses an incremental warm-start strategy to initialize quantum models with parameters from smaller trained quantum models, thus never initializing a model in a barren plateau. 

In the Red Cedar framework, models are trained via exact coordinate updates, in which each parameter is sequentially updated to that which reaches the minimum possible loss given the other parameter values. This method of training is guaranteed to converge to a local minimum of the loss function \cite{johri2025bit}. We find that we can further escape local optima in the model accuracy by using a loss function that is a weighted combination of the mean probability of measuring an incorrect output for a training sample and the variance of these probabilities. This method of training does not require a classical optimizer to update the parameters, unlike most quantum machine learning algorithms \cite{cerezo2021variational}. It also removes the need for setting optimizer hyperparameters, which model performance can be highly sensitive to \cite{moussa2024hyperparameter}. Due to the theoretical advantages over current quantum machine learning approaches, we employ the Red Cedar quantum classifier as the baseline quantum machine learning algorithm, and will refer to it as just the ``quantum model'' moving forward.

Using sub-net initialization, Red Cedar's incremental warm-start strategy, we start with a four qubit model and increase the number of qubits during training to six, eight, and then ten qubits. During training, colliding train samples are set to their majority label, and the entire training dataset is used as a batch to calculate the loss function. As the number of qubits increases, the number of collisions decreases, implying that the size of the batch increases. The number of qubits increases when a model reaches $100\%$ training accuracy on the current training data. For a given number of qubits and layers, we update all the parameters sequentially once. At the end of this, if we do not reach $100\%$ training accuracy, we add another layer of entangling nodes between each class qubit and feature qubit and continue training with the same number of qubits. We train the quantum model for up to five days and include the results that finish within that time frame.

For training, we use an ideal simulator, and do not explore the effects of noise. We confirm that shot noise is not a limiting factor in the final accuracy obtained --- reducing the number of shots reduces the training speed but not the final accuracy, in line with other studies \cite{adaptiveshot, qmlshot, variationalshotnoise}.

We emphasize that this training sequence is only one option. In particular, we stop model training when $100\%$ training accuracy is reached. However, we observe that the loss of samples in the test dataset can continue to decrease even after this point. The optimal stopping condition and sequence of updates are left to future work.

\subsection{Classical machine learning} \label{sec:cml}
While the quantum model is always trained on a bit-bit encoded dataset, we consider classical models trained on a dataset at different stages of data processing.

We define the stages of data processing as:
\begin{enumerate}
  \item \textit{Raw}: the dataset without any processing applied
  \item \textit{Preprocessed}: the dataset with preprocessing applied
  \item \textit{Pre-discretized bit-bit encoded}: the dataset with preprocessing and the data transformation steps of bit-bit encoding applied (shown in Figure~\ref{fig:fig1})
  \item \textit{Discretized bit-bit encoded}: the dataset with preprocessing and bit-bit encoding data transformations applied, discretized into some number of bits with information-based bit assignment (shown in Figure~\ref{fig:fig1})
\end{enumerate}

We train AutoML-optimized classical neural networks on the pre-discretized and discretized bit-bit encoded datasets in order to have baselines to compare the performance of the quantum model. We use the TPOT2 AutoML library \cite{ribeiro2024tpot2} to obtain neural networks with optimal architectures and hyperparameters for achieving maximal test accuracy while keeping runtime, energy usage, and manual configurations to a minimum. TPOT2, Tree-based Pipeline Optimization Tool 2, uses genetic programming to evolve a machine learning architecture that maximizes performance on a given dataset. TPOT2 works with scikit-learn \cite{pedregosa2011scikit} models, and here we constrain TPOT2 to only consider the MLPClassifier (neural network) for a fairer comparison to the quantum model. We allow the neural networks in the TPOT2 population to have the number of nodes in a layer at most be the number of features and at least be the number of classes. The neural networks can have between one and four hidden layers. We also allow the neural networks to have different solvers, learning rates, and L2 regularization terms. We use a population size of ten and we end the evolution process after at least ten generations if TPOT2's early stopping condition is reached, or after 20 minutes if not. We run TPOT2 on the training set only and use the one-vs-one area under the receiver operator curve as the fitness function.

We refer to the highest-performing model returned by TPOT2 as the AutoML-optimized classical neural network. For each preprocessing method, dataset, and train-test split, ten runs of TPOT2 are conducted on the bit-bit encoded dataset. The dataset is discretized into differing numbers of bits across the ten runs. The number of bits is set uniformly in the range $[1, Q_\text{dataset}(1.0) - \lceil \log_2\text{(number of classes)} \rceil]$. For example, if a dataset requires $20$ qubits allocated to the features to reach $100\%$ train and test accuracy, then ten discretized datasets are produced, with the first dataset encoded into two bits, the second dataset encoded into four bits, and so on until the last dataset which is encoded into $20$ bits. Then we run TPOT2 on each of those discretized datasets to obtain a classical AutoML-optimized neural network for each dataset. 

We conduct extra processing to the discretized bit-bit encoded datasets before training the classical neural networks. We split the binary representation of each feature such that each bit becomes its own feature. For example, a sample that had its feature vector encoded to $[2, 1]$ (\textit{i.e.} $[010, 1]$) would become $[0, 1, 0, 1]$. We then replace each $0$ with $-1$ for improved training capabilities.

\subsection{Statistics}
Error bars reported in both figures and text represent the standard error of the mean. All experiments are conducted across stratified five-fold train-test splits. We calculate statistical significance using the Mann-Whitney U test with a $95\%$ confidence level; comparisons reported as statistically significant have a $p \leq 0.05$, and comparisons reported as not significant have a $p > 0.05$.

\section{Results}
In order to evaluate the potential for quantum advantage in a diverse suite of oncological classification problems, we compare classification performance across optimized classical neural networks \cite{ribeiro2024tpot2}, the Red Cedar quantum machine learning framework \cite{johri2025bit}, and the theoretical maximum test accuracy any model could reach \cite{leither2025qubits}. We apply different, modality-appropriate preprocessing methods to the datasets and transform the data with bit-bit encoding \cite{johri2025bit} before model training.

Every quantum model that we analyze reaches $100\%$ training accuracy on its training dataset. This lets us focus our analysis on the performance on the test dataset, that is, on the generalizability of the model as a function of number of qubits. We emphasize that this is different from the aforementioned quantum machine learning studies which look at how training accuracy changes as a function of number of parameter updates. Since training accuracy reaches $100\%$, we are free to focus on the performance of the final model as a function of number of qubits.


\subsection{Wisconsin Diagnostic Breast Cancer} \label{sec:results:wdbc}
\begin{figure}
\centering
\includegraphics[width=\linewidth]{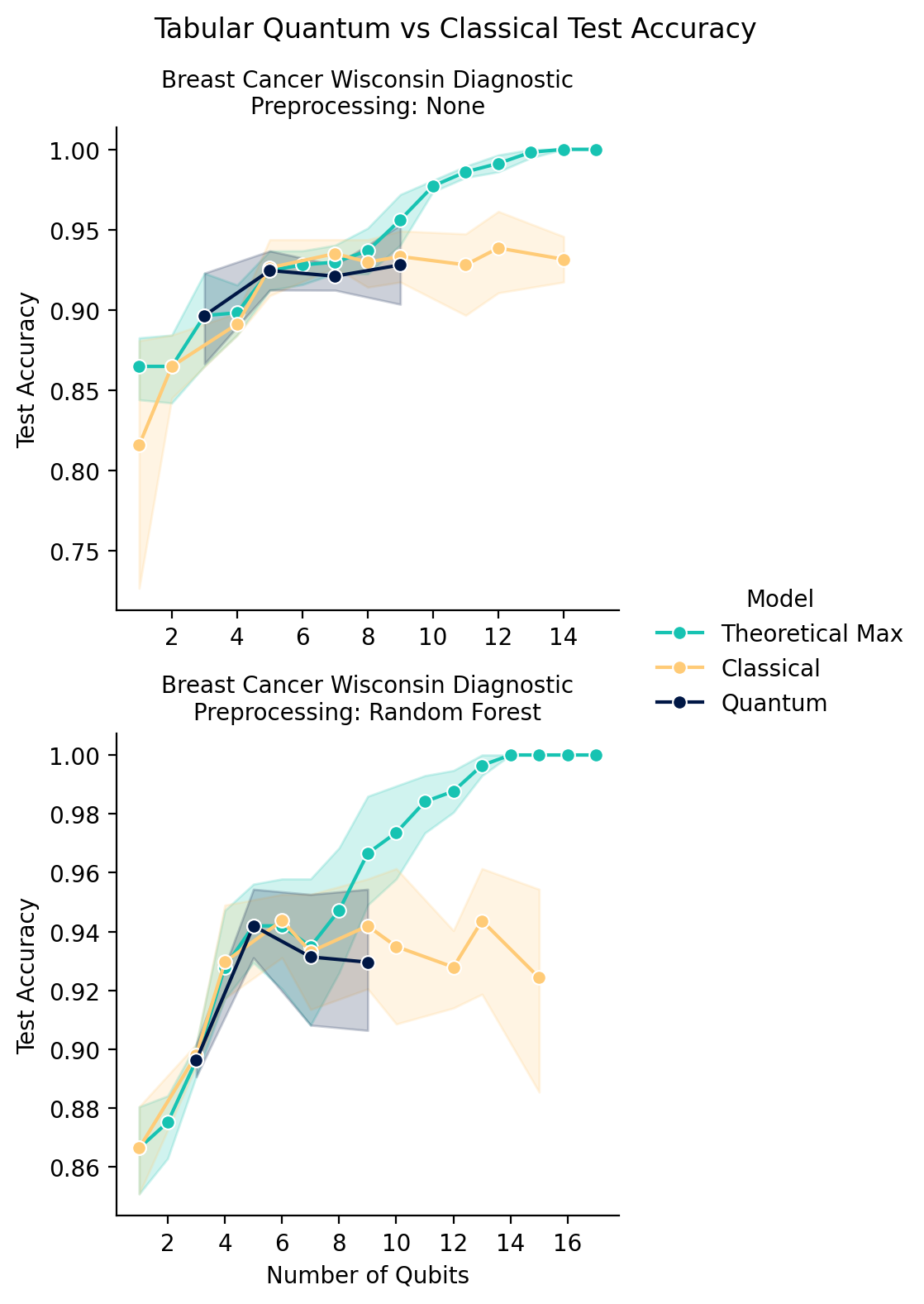}
\caption{The test accuracy and standard error of the theoretical, AutoML-optimized classical, and quantum models trained on the bit-bit encoded WDBC dataset with and without random forest preprocessing. The x-axis is the number of bits allocated to the dataset features. We see that on both the original and preprocessed dataset, the quantum and classical models fail to consistently reach the estimated maximum test accuracy and the quantum model performance is always within the standard error of the classical model performance.}
\label{fig:tabular}
\end{figure}

First, we investigate the WDBC dataset, which is commonly studied in both quantum and non-quantum machine learning literature (Section~\ref{sec:wdbc}). We explore both the original dataset and the dataset with random forest feature selection applied as a preprocessing method. Random forest feature selection reduces the dataset from $30$ features to $14 \pm 0.32$ with a classification train accuracy of $0.95 \pm 0.002$. We bit-bit encode both the original dataset and the dataset reduced to the features found to be important by random forest feature selection. The first step of bit-bit encoding is to transform the dataset with PCA dimensionality reduction to the number of components at which the cumulative explained variance reaches $90\%$, which transformed the original dataset to $7 \pm 0.0$ components and the preprocessed dataset to $5.2 \pm 0.37$. The statistically significant discrepancy between the number of components of the original and preprocessed dataset indicates that the removed features contributed novel variance to the dataset, which could be from classification-relevant information or noise.

The $Q_\text{dataset}(1.0)$, the number of qubits at which it becomes theoretically possible for both the train and test accuracy to reach $100\%$, provides insight into the potential for quantum advantage (Section~\ref{sec:re}). The original WDBC dataset's $Q_\text{dataset}(1.0) = 14.6 \pm 0.4$ and the preprocessed dataset's $Q_\text{dataset}(1.0) = 15.6 \pm 0.68$. A lack of statistical significance between the $Q_\text{dataset}(1.0)$s suggests that the differences in bit-bit encoding between the preprocessed and original datasets are not significant enough to lead to differences in downstream model classification accuracy. Neither of the $Q_\text{dataset}(1.0)$ values meets the $50$ qubit threshold indicating a potential for quantum advantage, which agrees with previous results indicating that benchmark datasets with less than $1,000$ features, less than $10,000$ samples, and less than ten classes do not appear to be likely candidates for quantum advantage \cite{leither2025qubits}.

While the resource estimation results suggest that WDBC is not a promising candidate for quantum advantage, we still investigate how quantum and classical model training performance behavior differs in this few-qubit regime. Figure~\ref{fig:tabular} shows the overlaid test accuracy of the theoretical, classical, and quantum models on the original and preprocessed bit-bit encoded WDBC datasets. We see the quantum model performance follows the trends of the classical model --- both models reach the theoretical maximum test accuracy at smaller numbers of bits, but fall below the theoretical maximum at larger numbers. The classical model and quantum model have different training methodologies regarding generalization, but for this dataset they reach statistically indistinguishable test accuracies for each number of qubits tested.

At some number of qubits, the average classical model test accuracy is higher than the theoretical maximum test accuracy. This is because the classical model breaks an assumption in the theoretical maximum test accuracy calculation that the training accuracy is at the theoretical maximum training accuracy for the allocated number of bits. The quantum model is specifically trained to reach the maximum theoretical train accuracy, while the AutoML that produces the final classical model attempts to balance train and test accuracy. Although the quantum train accuracy reaches the theoretical maximum train accuracy at each number of qubits, the quantum test accuracy does not reach the theoretical maximum test accuracy, because some test samples which do not overlap with train samples are misclassified, and the theoretical test accuracy calculation assumes that non-overlapping samples are classified correctly. While currently the quantum model trains until it reaches the theoretical maximum train accuracy, we plan to research other stopping points in training that better encourage generalizability.

\subsection{Omics}
\begin{figure*}
\centering
\includegraphics[width=\linewidth]{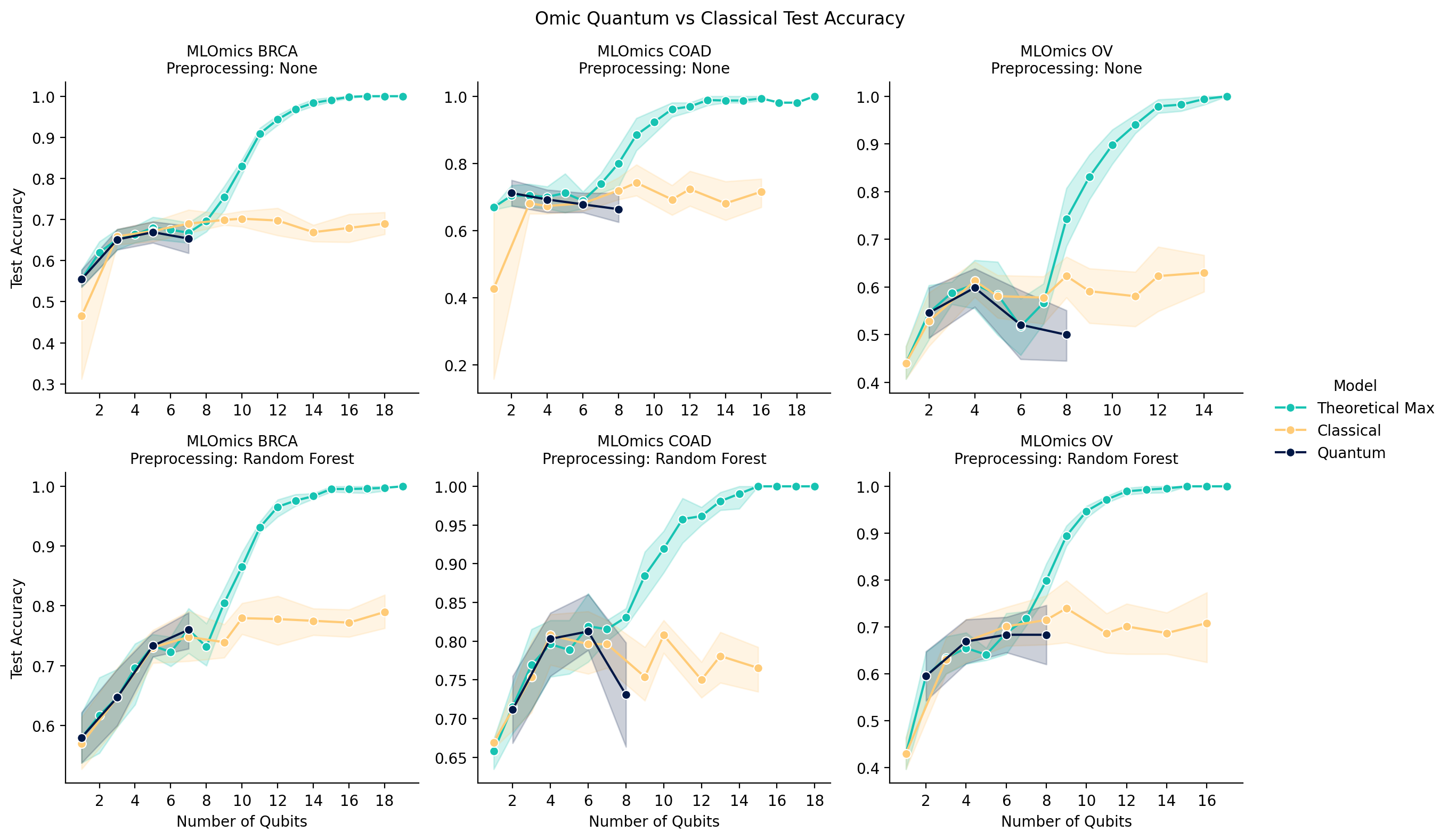}
\caption{The test accuracy of the theoretical, AutoML-optimized classical, and quantum models trained on the bit-bit encoded omics datasets (columns: BRCA, COAD, OV) without preprocessing (top row) and with random forest preprocessing (bottom row). In most cases, both models stop reaching the theoretical test accuracy at eight qubits, and the quantum model performance is within the standard error of the classical model performance. The classical models trained on the preprocessed datasets plateau at a higher accuracy than the classical models trained on the original datasets.}
\label{fig:omic}
\end{figure*}

We investigate the potential for quantum advantage on three different genomics datasets from TCGA / MLOmics: BRCA, COAD, and OV \cite{yang2025mlomics}. We explore how the potential for quantum advantage changes when utilizing original genomics datasets and genomic datasets with random forest feature selection preprocessing. Random forest feature selection reduces BRCA from $18206$ features to $43.8 \pm 1.068$ with a classification train accuracy of $0.84 \pm 0.005$, COAD from $17261$ features to $17.2 \pm 0.374$ with a classification train accuracy of $0.86 \pm 0.009$, and OV from $20684$ features to $21.4 \pm 0.400$ with a classification train accuracy of $0.84 \pm 0.009$. On average, random forest feature selection reduces the number of features in each omics dataset by $99.8\%$. This result concurs with previous work which has found that accurate classifications can be made on genomics datasets using only a small subset of genes \cite{wang2007accurate}. Without random forest preprocessing, bit-bit encoding reduces BRCA, COAD, and OV to $271.2 \pm 0.735$, $114.2 \pm 0.450$, and $131.0 \pm 0.0$ components, respectively. With preprocessing, bit-bit encoding reduces BRCA, COAD, and OV to $26.0 \pm 0.632$, $13.0 \pm 0.316$, and $14.4 \pm 0.400$ components, respectively.

The $Q_\text{dataset}(1.0)$ of the original BRCA, COAD, and OV omics datasets is $20.2 \pm 0.583$, $17.4 \pm 1.122$, and $15.8 \pm 0.583$. The preprocessed BRCA, COAD, and OV datasets have a $Q_\text{dataset}(1.0)$ of $20.6 \pm 0.510$, $16.8 \pm 0.860$, $17.2 \pm 0.735$. The lack of statistical significance between the original and preprocessed $Q_\text{dataset}(1.0)$s suggests that PCA successfully filters out classification-irrelevant noise from the top components for the omics datasets without the features being pre-filtered. The $Q_\text{dataset}(1.0)$s are all below the $50$ qubit threshold at which the quantum machine learning resource estimation framework suggests a potential for quantum advantage. The apparent lack of potential for quantum advantage is likely a side-effect of the small sample size of these datasets: BRCA has $671$ samples, COAD has $260$ samples, OV has $284$ samples. Previous work found that subsampling a genomic dataset with too few samples masks the true complexity of the dataset, as a dataset with too few samples will lack the representational power to exhibit the complexity of the true function mapping the features to the classes \cite{leither2025qubits}.

Figure~\ref{fig:omic} shows the overlaid test accuracy of the theoretical, classical, and quantum models on the original and preprocessed bit-bit encoded omics datasets. Similarly to Section~\ref{sec:results:wdbc}, we see the difference between the quantum and classical test accuracy is statistically insignificant for most number of qubits, across datasets and preprocessing methods. Also similarly to Section~\ref{sec:results:wdbc}, both models do not reach the theoretical maximum test accuracy as the number of qubits increase. Here, we also see the models struggle more to generalize when the omics datasets are not preprocessed --- the maximum test accuracy seen for both the classical and quantum models is significantly different between the original and preprocessed datasets, for each omics dataset. The models trained on the preprocessed datasets achieved higher test accuracy than the models trained on the original datasets, suggesting that random forest feature selection successfully reduces noise in the datasets that affect the quality of the PCA components. Yet, the $Q_\text{dataset}(1.0)$ is not significantly different between the original and preprocessed datasets. Above $\log_2(\text{number of samples})$ bits, the number of unique representations possible exceeds the total number of samples, which may lead to redundancy in the discretized dataset that makes the $Q_\text{dataset}(1.0)$ metric robust against noise.

\subsection{Spatial}
\begin{figure*}
\centering
\includegraphics[width=\linewidth]{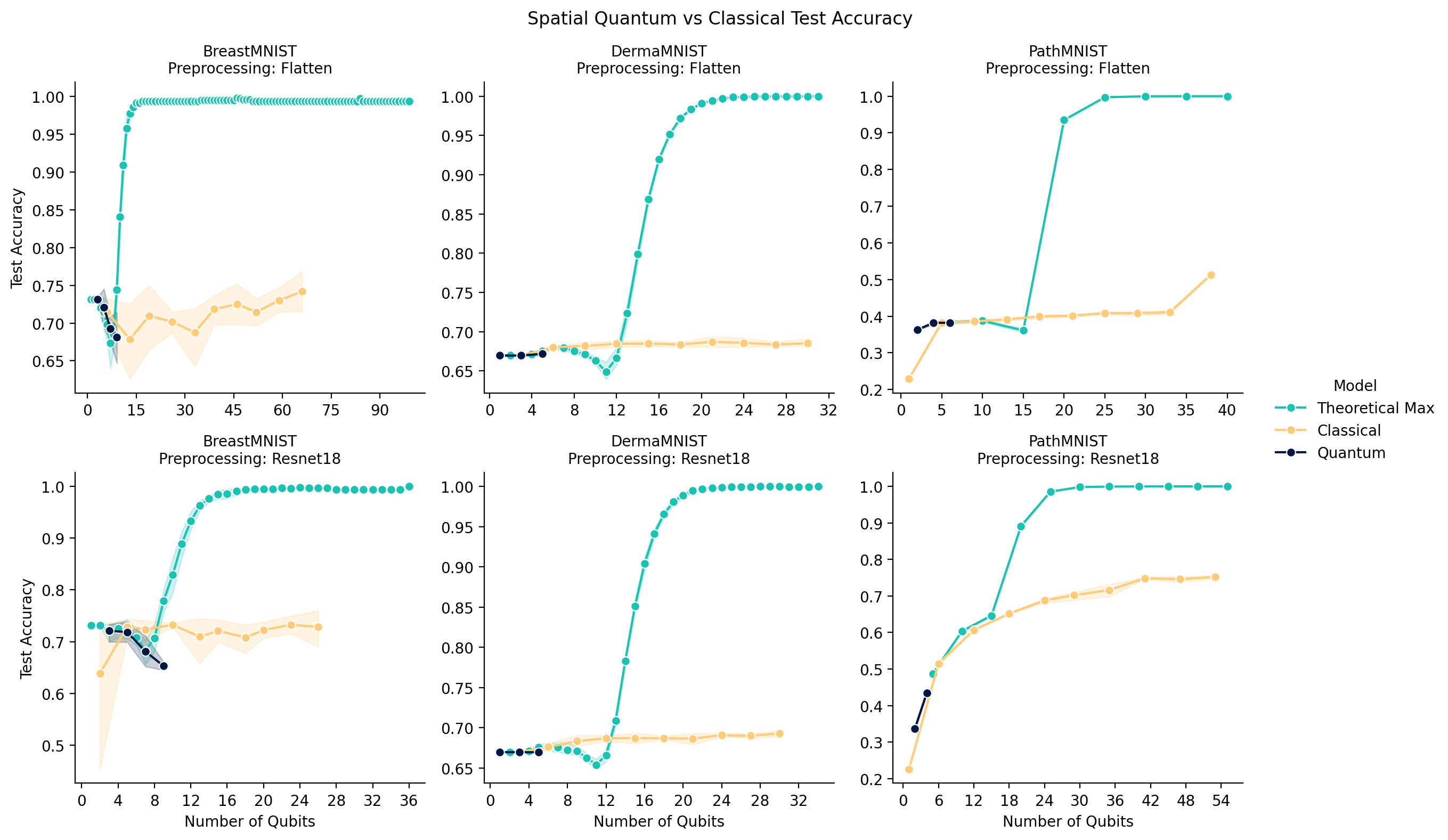}
\caption{The test accuracy of the theoretical, AutoML-optimized classical, and quantum models trained on the bit-bit encoded spatial datasets (columns: BreastMNIST, DermaMNIST, PathMNIST) with flattening (top row) and with ResNet18 preprocessing (bottom row). The theoretical maximum test accuracy reaches $100\%$ at a much larger number of qubits than the previous datasets. The spatial datasets exhibit unique behaviors in test accuracy differences between models trained on the flattened datasets compared to the models trained on the ResNet18 preprocessed datasets.}
\label{fig:spatial}
\end{figure*}

We investigate the potential for quantum advantage on three different oncological spatial datasets from MedMNIST: BreastMNIST, DermaMNIST, and  PathMNIST \cite{yang2023medmnist}. We explore how the potential for quantum advantage changes when using flattened image pixel values and using the output vector from passing images through ResNet18. Since MedMNIST standardizes image sizes to $28$x$28$, each flattened dataset has $28\cdot28\cdot3 = 2352$ features and each ResNet18 preprocessed dataset has $512$ features. With flattening, bit-bit encoding reduces BreastMNIST, DermaMNIST, and PathMNIST to $135.4 \pm 0.400$, $16.8 \pm 0.200$, and $178.4 \pm 0.245$ components. With ResNet18 preprocessing, bit-bit encoding reduces BreastMNIST, DermaMNIST, and PathMNIST to $130.6 \pm 0.400$, $174.0 \pm 0.0$, and $179.0 \pm 0.0$ components. While the differences between the retained components for the flattened and ResNet18 preprocessed BreastMNIST and PathMNIST datasets are not significant, DermaMNIST has $88\%$ fewer kept components when flattened than when preprocessed with ResNet18. This suggests that DermaMNIST is more compressible than the other spatial datasets and that ResNet18 may not capture DermaMNIST well.

The $Q_\text{dataset}(1.0)$s of the flattened BreastMNIST, DermaMNIST, and  PathMNIST are $66.2 \pm 11.015$, $32.2 \pm 0.800$, and $42.0 \pm 1.225$. The ResNet18 preprocessed BreastMNIST, DermaMNIST, and  PathMNIST $Q_\text{dataset}(1.0)$s are $26.8 \pm 2.8$, $32.8 \pm 1.2$, $56.5 \pm 1.443$. The high stochasticity in the $Q_\text{dataset}(1.0)$ for flattened BreastMNIST suggests the presence of outlier samples that affect how easily representable the dataset is depending on how the outliers are distributed across the train and test splits. The drop in $Q_\text{dataset}(1.0)$ for ResNet18 preprocessed BreastMNIST then suggests that ResNet18 successfully filters out noise during dataset compression. Given the stochasticity and difference in $Q_\text{dataset}(1.0)$ between preprocessing methods, along with the high number of components kept after PCA, the high $Q_\text{dataset}(1.0)$ for flattened BreastMNIST suggests poor processing of the dataset rather than a true potential for quantum advantage. The other two datasets border on the quantum advantage 50-qubit threshold which warrants further investigation in future work.

Figure~\ref{fig:spatial} shows the overlaid test accuracy of the theoretical, classical, and quantum models on the flattened and ResNet18 preprocessed bit-bit encoded spatial datasets. Note that some quantum models did not finish running within the time frame so higher-qubit performances are not available. The dip in the theoretical accuracy, which is not seen in the other datasets, suggests that there may be more optimal ways to encode such data to be explored in future work. As discussed in Section~\ref{sec:results:wdbc}, the classical models can achieve higher accuracy than the theoretical maximum test accuracy because the classical model train accuracy does not reach the theoretical maximum train accuracy. We see that ResNet18 preprocessing makes minimal difference to the generalizability of the models trained on the BreastMNIST and DermaMNIST datasets. For PathMNIST, ResNet18 preprocessing leads to a significant increase in the classical model test accuracies compared to the classical models trained on the flattened PathMNIST datasets, despite the $Q_\text{dataset}(1.0)$ being lower for the flattened PathMNIST than the ResNet18 preprocessed PathMNIST.

Furthermore, the classical model test accuracy appears to be trending slightly upward as the number of bits allocated to the dataset increases. We train the classical model only up to the average number of feature bits at which a model can theoretically achieve $100\%$ accuracy. The upward trend in accuracy suggests that increased performance may be seen if more bits are allocated to the dataset, which could then suggest that more redundancy in classification-relevant information is necessary for classical model generalization. The quantum model accuracy follows the theoretical test accuracy, but since we only ran the quantum model up to ten qubits and the spatial datasets have much larger $Q_\text{dataset}(1.0)$s, it is difficult to make broader observations.

\section{Discussion} \label{sec:discussion}
In this paper, we investigated the potential for quantum advantage in oncological classification problems across a range of data modalities and preprocessing methods. We found that none of the tested learning problems exhibited strong evidence of a potential for quantum advantage. The WDBC and omics datasets fall well below the 50-qubit threshold for quantum advantage candidacy; some spatial datasets approach or exceed the threshold, but did not consistently do so across preprocessing methods. Additionally, up to the model sizes trained, the quantum and classical models have similar testing accuracy. We emphasize that since the quantum models reach $100\%$ training accuracy, this outcome is unlikely to be an artifact of the particular training methodology and model architecture used here. 

Given that the datasets explored here are all common benchmark datasets seen in other works attempting to demonstrate quantum advantage, our results suggest that quantum machine learning researchers should explore higher-dimensional datasets more likely to exhibit the representational power and higher-order interactions that are thought to lead to quantum advantage. Due to the sensitive nature of oncological and biological data, there is currently a lack of easily accessible open-source datasets with the apparent required complexity. Collaborations with the healthcare and biotech industries, which have access to multitudes of real-world data, could be fruitful for demonstrating the utility of quantum machine learning.

We found that our quantum machine learning resource estimation method is robust to different methods of preprocessing a dataset. We saw this particularly with the omics datasets --- while preprocessing the omics datasets reduced the number of features by $99.8\%$ on average, the $Q_\text{dataset}(1.0)$ between the non-preprocessed and preprocessed datasets was statistically insignificant. This result suggests that the quantum machine learning resource estimation framework can provide estimates about the potential for quantum advantage without being too reliant on how well a researcher managed to clean and compress the data beforehand. While our study found the quantum machine learning resource estimation framework to be robust to preprocessing method, previous results found that the framework is sensitive to the specifics of the parameters within bit-bit encoding \cite{leither2025qubits}. Choosing the proper dimensionality reduction scheme, number of features post dimensionality reduction, and bit allocation algorithm is still important --- a large presence of class-irrelevant information in encoding can result in inflated quantum advantage estimates.

Given the early state of research with the quantum machine learning resource estimation framework, there is a range of exciting directions to take this and related work in the future that also could help address some limitations in this paper. All of the actual quantum model results were run on a simulator. Running on a simulator restricts the depth of analyses possible to make when comparing the quantum model performance to the classical model performance; for example, on quantum hardware we could investigate the impact of decoherence as we increase the number of logical qubits. While we used AutoML-optimized classical models to attempt to have the best classical model performance possible for the fairest comparisons, is possible that the AutoML algorithm did not find a model that could achieve the best possible performance on a given dataset. This is why having the quantum machine learning resource estimation framework to provide an estimation of the potential for quantum advantage without needing to rely purely on model comparisons is important. 

Finally, we limited our experiments to a tabular dataset, single-omics data, and spatial data. Investigating the potential for quantum advantage on rich datasets such as multi-omics or multi-modal, or different problem types such as time-series, regression, or multi-label problems may yield more promising results even on open-source benchmark datasets. Many of these dataset types require or are better served with model architectures beyond classification neural networks --- a framework for quantum advantage estimations meant specifically for different kinds of data may lead to more informative estimations. For now, our approach of comparing a quantum model to AutoML-optimized classical models across scales, and supplementing the comparisons with the quantum machine learning resource estimation framework, provides the first step towards a more robust approach for finding quantum machine learning advantage.


\section{Software Framework}
The techniques in the paper are implemented using Red Cedar, a commercial software framework for quantum machine learning being developed at Cascade Quantum, Inc. It can be made available upon request. The publicly available Red Cedar resource estimation framework used to generate the theoretical estimates in the figures can be accessed at \href{qre.cascadequantum.com}{qre.cascadequantum.com}.

\bibliographystyle{IEEEtran}
\bibliography{biblio}

\end{document}